\documentclass[10pt]{article}
\usepackage{latexsym}
\usepackage{amsfonts}
\usepackage{graphicx}
\usepackage{epsfig}
\usepackage{amsmath}
\usepackage{amssymb}
\usepackage{mathrsfs}
\usepackage{color}

\begin{document}
\bibliographystyle{unsrt} 
\newtheorem{theorem}{Theorem}[section]
\newtheorem{lemma}[theorem]{Lemma}
\newtheorem{corollary}[theorem]{Corollary}
\newtheorem{definition}[theorem]{Definition}
\newtheorem{remark}[theorem]{Remark}

\def\proof{{\noindent{\sc Proof.\ }}}
\newcommand{\QED}{\hspace{1ex}\hfill$\Box$\vspace{2ex}}

\newcommand{\eps}{\varepsilon}
\def\R{\mathbb{R}}
\def\C{\mathbb{C}}

\newfont{\tenbss}{bbmss10}
\newcommand{\bsa}{\mbox{\tenbss a}}
\newcommand{\bsc}{\mbox{\tenbss c}}
\newcommand{\bsf}{\mbox{\tenbss f}}
\newcommand{\bsz}{\mbox{\tenbss z}}

\newfont{\sforf}{cmss10}
\newfont{\tenss}{lcmss8 at 10pt}
\newfont{\ltenss}{lcmss8}
\newcommand{\saa}{\mbox{\tenss a}}       
\newcommand{\sse}{\mbox{\tenss e}}
\newcommand{\ssf}{\mbox{\sforf f}}

\def\intav{{-}\hspace{-2.5ex}\int}
\def\intavm{\mathop{\int\hspace{-2.1ex}{\vspace{-0.5ex}-}}}
\newcommand\LL{\hbox to 7pt{\vrule width.4pt \vbox to 7pt{\vfill \hrule width 5pt height.4pt}}}
\def\weak{\rightharpoonup}
\def\weakstar{\stackrel{*}{\rightharpoonup}}

\newcommand\calA{\mathcal{A}}
\newcommand\calF{\mathcal{F}}
\newcommand\calG{\mathcal{G}}
\newcommand\calH{\mathcal{H}}
\newcommand\calK{\mathcal{K}}
\newcommand\calL{\mathcal{L}}
\newcommand\calM{\mathcal{M}}
\newcommand\calR{\mathcal{R}}
\newcommand\calS{\mathcal{S}}
\newcommand\calW{\mathcal{W}}

\newcommand{\Om}{\Omega}
\newcommand{\pOm}{\partial\Omega}
\newcommand{\pDOm}{\partial_D\Omega}
\newcommand{\pNOm}{\partial_N\Omega}

\newcommand{\Ome}{\Omega^\eps}
\newcommand{\pOme}{\partial\Omega^\eps}
\newcommand{\pDOme}{\partial_D\Omega^\eps}
\newcommand{\pNOme}{\partial_N\Omega^\eps}

\newcommand{\om}{\omega}
\newcommand{\pom}{\partial\omega}
\newcommand{\pDom}{\partial_D\omega}
\newcommand{\pNom}{\partial_N\omega}

\renewcommand{\div}{\text{div}\,}
\newcommand{\Hdive}{H({\rm div},\Omega_\eps)}
\newcommand{\Hdiv}{H({\rm div},\Omega)}

\title{\bf Plate theory as the variational limit\\ of the complementary energy functionals \\ of inhomogeneous anisotropic\\ linearly elastic bodies }

\author{{\sc Fran\c{c}ois Murat}\thanks{Laboratoire Jacques-Louis Lions,
Universit\'{e} Pierre et Marie Curie (UPMC Paris VI), Bo\^{i}te courrier 187, 75252 Paris Cedex
05, France,
       email: {\tt murat@ann.jussieu.fr}}\and
{\sc Roberto Paroni}\thanks{DADU,  Universit\`{a} degli Studi di Sassari,
Palazzo del Pou Salit, Piazza Duomo 6, 07041 Alghero, Italy, email:
{\tt paroni@uniss.it} } }

\maketitle

\begin{abstract}
We consider a sequence of linear hyper-elastic, inhomogeneous and fully anisotropic bodies in a reference configuration occupying a cylindrical region of height $\eps$. We then study, by means of
$\Gamma$-convergence, the asymptotic behavior as $\eps$ goes to zero of the sequence of complementary energies.  The limit functional is then identified as a dual problem for a two-dimensional plate.
Our approach gives a direct characterization of the convergence of the equilibrating stress fields.
 \end{abstract}

\vspace{2ex}

\noindent{\small{\it 2010 Mathematics Subject Classification}:
49S05, 49J45,  74K20, 74B05, 

\vspace{2ex}

\section{Introduction}

The equilibrium problem for a linear hyper-elastic body may be suitably studied by means of several variational formulations, like the principle of the minimum potential energy (primal formulation) and the principle of minimum complementary energy (dual formulation). In the former formulation the unknown is the displacement vector field, while in the latter the stress tensor field is to be found. Other variational formulations, the so called mixed formulations, take simultaneously as unknowns the displacement and the stress vector fields, see for instance \cite{Gu1972}.

In the last three decades, starting with the work of Ciarlet and Destuynder \cite{CD1979}, these variational problems, or their extremality equations, have been used, in conjunction to some asymptotic techniques, to justify/derive models for thin structures starting from the three-dimensional theory.
At the early stages of this prolific line of research mixed formulations were adopted, while after the asymptotic techniques have been refined the research have been focused almost exclusively on the study of some form of the primal formulation.
Within this line of research,
the Kirchhoff-Love theory for homogeneous and isotropic plates has been justified by means of $\Gamma$-convergence by
Anzellotti {\it et al.} \cite{ABP94} and by Bourqin {\it et al.} \cite{BCGR92}. These results have been generalized in several directions: for linear plates with residual stress \cite{Pa06d,Pa06s}, for elasto-plastic plates \cite{LR12, DM13, Da14}, Reissner-Mindlin plates \cite{PPGT06, PPGT07, NHJ10}, and non-linearly elastic plates \cite{FJM02, FJM06}. 

Respect to the existing literature a different route has been taken by  Bessoud {\it et al.} in \cite{BGK12}. These authors consider a system of two elastic materials glued by a thin and strong material between them and by means of the complementary energy they study the asymptotic behavior
of the system of materials  as the thickness of the gluing material goes to zero. In the  limit problem a material surface, endowed with an appropriate elastic energy, replaces the thin layer.

We here consider a sequence of linear hyper-elastic, inhomogeneous, and fully anisotropic bodies in a reference configuration occupying a cylindrical region of height $\eps$. We then study, by means of
$\Gamma$-convergence, the asymptotic behavior as $\eps$ goes to zero of the sequence of complementary energies.  The limit functional is then identified as a dual problem for a two-dimensional plate. 

While variational limits of primal problems characterize the asymptotic behavior of the minimizing displacements, the study of the asymptotic behavior of the complementary energies  characterizes the convergence of the equilibrating stress fields. Besides the use of this novel approach for the deduction of plate theory, our work deals with fully anisotropic and inhomogeneous materials, case that has not been studied in this full generality before, not even by means of the primal formulation. This kind of generality on the constitutive equations has been used to derive linearly elastic beam theories, within the primal formulation framework, for instance in \cite{MS95, MS99, MS00, FMP08}.

The paper is organized as follows. In Section \ref{sec2} we review some function spaces that will be useful in the rest of the paper, while in Section \ref{sec3} the primal and dual formulation of the problem considered are stated. The dimension reduction problems are classically  rescaled on a fixed domain, this is done in Section \ref{sec4}. In Section \ref{sec5} the $\Gamma$-convergence analysis is carried on, and in Section \ref{sec6} the obtained $\Gamma$-limit is written on a two-dimensional domain.

\section{Preliminaries}\label{sec2}

Let $\Omega\subset\R^3$ be an open, bounded set with Lipschitz boundary $\pOm$, and let $\Gamma$ be an open subset of $\pOm$.\footnote{Note that no regularity assumption is made on the open set $\Gamma$.} 
We denote by 
$$
H^{1/2}(\Gamma;\R^3):=\{v:\exists u\in H^1(\Omega;\R^3) \mbox{ s.t. }\gamma u=v \mbox{ on }\Gamma\},
$$ 
where $\gamma: H^1(\Omega;\R^3)\to H^{1/2}(\pOm;\R^3) $ denotes the trace operator, and we equip it with the norm
$$
\|v\|_{H^{1/2}(\Gamma)}:=\inf\{\|u\|_{H^1(\Omega)}:u\in H^1(\Omega;\R^3) \mbox{ and }\gamma u=v \mbox{ on }\Gamma\}.
$$
The dual of $H^{1/2}(\Gamma;\R^3)$ shall be denoted by $H^{-1/2}(\Gamma;\R^3)$.
We let
$$
H^{1/2}_{00}(\Gamma;\R^3):=\{v\in H^{1/2}(\Gamma;\R^3):\tilde v\in H^{1/2}(\pOm;\R^3)\},
$$
where $\tilde v$ denotes the extension by $0$ of $v$ to $\pOm$, and we equip it with the norm
$$
\|v\|_{H^{1/2}_{00}(\Gamma)}:=\|\tilde v\|_{H^{1/2}(\pOm)}.
$$
The spaces $H^{1/2}(\Gamma;\R^3)$ and $H^{1/2}_{00}(\Gamma;\R^3)$ are delicate spaces, see e.g., see \cite{LM1972} Chapter I, \S 11 and 12, and \cite{Gr1985} Chapter 1, \S 1.3.2. Note that the space denoted here, and in \cite{LM1972}, by $H^{1/2}_{00}(\Gamma)$ is denoted by $\tilde W^{1/2}_{2}(\Gamma)$ in \cite{Gr1985}.

Thanks to these spaces, for a distribution $f\in H^{-1/2}(\pOm;\R^3)$ defined in the whole boundary $\pOm$, we may define its restriction to $\Gamma$, denoted by $f|_\Gamma\in (H^{1/2}_{00}(\Gamma;\R^3))'$, in the following way
$$
\langle f|_\Gamma,v\rangle_{H^{1/2}_{00}(\Gamma)}:=\langle f,\tilde v\rangle_{H^{1/2}(\pOm)}\quad
\mbox{for every }v \in H^{1/2}_{00}(\Gamma;\R^3).
$$
 
The space
$$
\Hdiv:=\{T\in  L^2(\Om;\R^{3\times3}_{\rm sym}):\div T\in  L^2(\Om;\R^{3})\},
$$
equipped with the norm
$$
\|T\|_{\Hdiv}^2:=\|T\|^2_{L^2(\Om)}+\|\div T\|^2_{L^2(\Om)},
$$
is a Hilbert space.
It is well known      that there exists a continuous linear mapping $\gamma_n:\Hdiv\to H^{-1/2}(\pOm;\R^3)$ such that
\begin{equation}\label{ggg}
\int_\Om T\cdot\nabla u\,dx=-\int_\Om \div T\cdot u\,dx + \langle \gamma_n T,\gamma u\rangle_{H^{1/2}(\pOm)},
\end{equation}
for every $T\in \Hdiv$ and $u\in H^1(\Om;\R^3)$.
Hereafter we shall simply write $Tn$ in place of $\gamma_n T$.

From \eqref{ggg} it follows that  for
$T\in \Hdiv$ and for every $u\in H^1(\Om;\R^3)$ with $\gamma u=0$ in $H^{1/2}(\pOm\setminus\Gamma;\R^3)$
we have
\begin{equation}\label{pgg}
\int_\Om T\cdot\nabla u\,dx=-\int_\Om \div T\cdot u\,dx + \langle Tn|_\Gamma,\gamma u\rangle_{H^{1/2}_{00}(\Gamma)},
\end{equation}
since $\gamma u\in H^{1/2}_{00}(\Gamma;\R^3)$.

Hereafter, if no confusion shall arise, we shall simply write $Tn$ also for the restriction $Tn|_\Gamma$ and we shall drop the use of $\gamma$ to denote the trace, i.e., we shall write $u$ for the trace $\gamma u$.

\section{The unscaled problems}\label{sec3}

Let $\omega$ be an open bounded domain of $\R^2$ with Lipschitz boundary $\partial\omega$, and for $\eps\in (0,1]$, we set
$$\Ome:=\omega\times (-\eps/2,\eps/2).$$
Let $\pDom$ and $\pNom$ be unions of finite numbers of open connected subsets of $\partial\omega$  such that
$$\pDom\cap\pNom=\emptyset, \quad \overline{\pDom}\cup\overline{\pNom}=\pom,\quad \pDom\ne \emptyset.$$

We set
 $$
 \pDOme:= \pDom\times (-\eps/2,\eps/2), \quad
 \pNOme:=\pOme\setminus\overline{\pDOme}.$$
 We consider  $\Ome$ as the region occupied by a linear hyper-elastic body in the reference configuration. Let  $\hat \C^\eps \in L^\infty(\Ome;\R^{3\times3\times3\times3}_{\rm sym})$
 be the elasticity tensor, which we assume to be uniformly coercive, of the elastic body considered.
 By writing  $\hat \C^\eps \in L^\infty(\Ome;\R^{3\times3\times3\times3}_{\rm sym})$ we mean that
 $$
 \hat \C^\eps_{ijkl}=\hat \C^\eps_{klij}=\hat \C^\eps_{ijlk}.
 $$ 
The sets $\pDOme$ and $\pNOme$ are the parts of the boundary of $\Ome$ where Dirichlet and Neumann boundary conditions are imposed, and we denote\footnote{
Throughout the paper the notation $\hat \cdot$ refers to quantities defined on $\Ome$ or on parts of its boundary.} by $\hat f^\eps\in(H^{1/2}_{00}(\pNOme;\R^3))'$ the surface loads, and by $\hat g^\eps \in H^{1/2}(\pDOme;\R^3)$ the imposed displacement on $\pDOme$. Since every function in $H^{1/2}(\pDOme;\R^3)$ is the trace of a function in $H^1(\Ome;\R^3)$, we also denote by $\hat g^\eps$ this latter function.

We further denote by $\hat b^\eps\in L^{2}(\Ome;\R^3)$ the body forces.
\begin{remark}\label{remf}
Let $\omega_\pm^\eps$ be the upper and lower bases of the cylinder $\Ome$ and let
$ \pNOme_\ell$ be the Neumann part of the lateral boundary, i.e.,
$$
 \omega_\pm^\eps:=\omega\times\{\pm{\eps}/{2}\}, \quad \mbox{and}\quad 
 \pNOme_\ell:=\pNom\times (-{\eps}/{2},+{\eps}/{2}),
$$
so that
$
\omega_+^\eps\cup \omega_-^\eps\cup\pNOme_\ell$
is $\pNOme$ up to a set of zero two-dimensional measure.
 Let $\hat {\bar f}_\pm^\eps\in H^{-1/2}(\om^\eps_\pm; \R^3)$, and  $\hat {\bar f}_\ell^\eps\in H^{-1/2}(\pNOme_\ell;\R^3)$.
Then, $\hat {\bar f}^\eps$ defined\footnote{
Throughout the paper the notation $\hat {\bar \cdot}$ refers to quantities which are examples of the general case.}, for every $\hat v \in H^{1/2}_{00}(\pNOme;\R^3)$, by
\begin{align}
\langle \hat {\bar f}^\eps,\hat v\rangle_{H^{1/2}_{00}(\pNOme)}:=&\langle \hat {\bar f}_+^\eps,\hat v|_{\omega_+^\eps}\rangle_{H^{1/2}(\om_+^\eps)}+\langle \hat {\bar f}_-^\eps,\hat v|_{\omega_-^\eps}\rangle_{H^{1/2}(\om_-^\eps)}\nonumber\\
&\hspace{0cm}+\langle \hat {\bar f}_\ell^\eps,\hat v|_{\pNOme_\ell}\rangle_{H^{1/2}(\pNOme_\ell)},\label{36}
\end{align}
 is an example of a force that can be used as $\hat f^\eps$. Moreover, in the case that $\hat {\bar f}_\pm^\eps\in L^{2}(\om^\eps_\pm; \R^3)$ and  $\hat {\bar f}_\ell^\eps\in L^{2}(\pNOme_\ell;\R^3)$, the duality in \eqref{36} is nothing but the sum of three integrals. Note however that there are forces in $(H^{1/2}_{00}(\pNOme;\R^3))'$ that are more general than $\hat{\bar f}^\eps$ defined by \eqref{36}.
\end{remark}
\begin{remark}\label{GLe}
Given $\hat f^\eps\in(H^{1/2}_{00}(\pNOme;\R^3))'$ and $\hat b^\eps\in L^{2}(\Ome;\R^3)$, as above,
there exists (looking e.g.\ for $\hat G^\eps$ as the symmetric part of the gradient of an unknown function) a $\hat {G}^\eps \in \Hdive$
such that
\begin{eqnarray}\label{eqFe}
\left\{
\begin{array}{ll}
\mbox{\rm div}\, \hat {G}^\eps+\hat b^\eps=0 & \mbox{ in }L^2(\Ome;\R^3),\\
\hat {G}^\eps \hat n=\hat f^\eps & \mbox{ in }(H^{1/2}_{00}(\pNOme))'.
\end{array}
\right.
\end{eqnarray}
Then, from \eqref{pgg}, the work done by the loads can be simply  rewritten as
 \begin{equation}\label{ggF}
\int_{\Ome} \hat b^\eps\cdot \hat v\,d\hat x+
\langle \hat f^\eps,\hat v\rangle_{H^{1/2}_{00}(\pNOme)}=\int_{\Ome}\hat {G}^\eps\cdot E\hat v\,d\hat x,
\end{equation}
for all $\hat v\in H^1(\Ome;\R^3)$ such that $\hat v =0$ on $\pDOme$.

Also the converse is true: given $\hat {G}^\eps \in \Hdive$  there exist  $\hat f^\eps\in(H^{1/2}_{00}(\pNOme;\R^3))'$ and $\hat b^\eps\in L^2(\Ome;\R^3)$, defined by \eqref{eqFe}, for which \eqref{ggF} holds.

Therefore the description of the applied loads may be done indifferently either by means of the body and surface forces $\hat b^\eps$ and $\hat f^\eps$, or by means of the tensor field $\hat {G}^\eps$. 
Both approaches present some advantages and some disadvantages. For instance, it is not necessary to assume  $\hat {G}^\eps \in \Hdive$ but it is enough to have
$\hat {G}^\eps \in L^2(\Ome;\R^{3\times 3}_{\rm sym})$. 
We  consider hereafter both representations simultaneously. In Remark \ref{GL}, below, we explain why we consider both type of forces.
\end{remark}

We consider, in the spirit of  Remark \ref{GLe}, also ``generalized forces" described by a tensor field $\hat H^\eps \in L^2(\Ome;\R^{3\times 3}_{\rm sym})$. 

The problem of linear elasticity can be written as:
\begin{equation}\label{pbleeps0}
\left\{\begin{array}{l}
\hat w^\eps\in H^1(\Ome;\R^3),\hat w^\eps=\hat g^\eps \mbox{ in }H^{1/2}(\pDOme;\R^3),\\
\displaystyle\int_{\Ome}\hat\C^{\eps}E\hat w^\eps\cdot E\hat\psi\,d\hat x=
\int_{\Ome}\hat H^\eps\cdot E\hat \psi +
\hat b^\eps\cdot \hat \psi\,d\hat x+\langle \hat f^\eps,\hat \psi\rangle_{H^{1/2}_{00}(\pNOme)},\\
\mbox{for every } \hat\psi\in\hat \calA^\eps,
\end{array}
\right.
\end{equation}
where $E\hat w^\eps$ denotes the symmetric part of the gradient of $\hat w^\eps$, and
$\hat \calA^\eps$ the set of admissible displacements defined by
$$
\hat \calA^\eps:=\{\hat v\in H^1(\Ome;\R^3):\hat v =0 \mbox{ on }\pDOme\}.
$$

Since $\hat {g}^\eps$ simultaneously denotes a function in $H^1(\Ome;\R^3)$ and its trace, \eqref{pbleeps0} can be rewritten as:
\begin{equation}\label{pbleeps}
\left\{\begin{array}{l}
\hat u^\eps:=\hat w^\eps-\hat {g}^\eps\in \hat \calA^\eps,\\
\displaystyle\int_{\Ome}\hat\C^{\eps}E\hat u^\eps\cdot E\hat\psi\,d\hat x=
\int_{\Ome}\hat F^\eps\cdot E\hat \psi +
\hat b^\eps\cdot \hat \psi\,d\hat x+\langle \hat f^\eps,\hat \psi\rangle_{H^{1/2}_{00}(\pNOme)},\\
\mbox{for every } \hat\psi\in\hat  \calA^\eps,
\end{array}
\right.
\end{equation}
where, for notational simplicity, we denote 
\begin{equation}\label{39}
\hat F^\eps:=\hat H^\eps+\hat\C^{\eps}E\hat {g}^\eps.
\end{equation}

As it is well known, the solution $\hat u^\eps$ of \eqref{pbleeps} may also be found by minimizing the total energy  $\hat \calF^\eps:\hat \calA^\eps \to\R$ 
defined by
$$
\hat \calF^\eps(\hat v):=\frac 12 \int_{\Ome}\hat\C^\eps E\hat v\cdot E\hat v\,d\hat x-\int_{\Ome}
\hat F^\eps\cdot E\hat v +
\hat b^\eps\cdot \hat v\,d\hat x-\langle \hat f^\eps,\hat v\rangle_{H^{1/2}_{00}(\pNOme)},
$$
that is 
$$
\hat \calF^\eps(\hat u^\eps)=\inf_{\hat v\in \hat \calA^\eps}\hat \calF^\eps(\hat v).
$$
This variational problem is called \textsc{Primal Problem}.

We now introduce the dual problem.

From the inequality
$$
0\le \hat\C^\eps (\hat E- (\hat\C^\eps)^{-1}\hat S)\cdot (\hat E- (\hat\C^\eps)^{-1}\hat S)= 
\hat \C^\eps \hat E\cdot \hat E-2\hat E\cdot \hat S + (\hat\C^\eps)^{-1}\hat S\cdot \hat S,
$$
which holds for every $\hat E, \hat S\in\R^{3\times3}_{\rm sym}$,
it follows that for every $\hat v\in H^1(\Ome;\R^3)$
$$
\frac 12 \int_{\Ome}\hat\C^\eps E\hat v\cdot E\hat v\,d\hat x=\max_{\hat S\in L^2(\Ome;\R^{3\times3}_{\rm sym})} \int_{\Ome}\hat S\cdot E\hat v- \frac 12(\hat\C^\eps)^{-1}\hat S\cdot\hat S\,d\hat x
$$
with the max achieved for $\hat S=\hat\C^\eps E\hat v$.
Thus we can rewrite the direct problem as
$$
\inf_{\hat v\in \hat \calA^\eps}\max_{\hat S\in L^2(\Ome;\R^{3\times3}_{\rm sym})}\hat \calL^\eps(\hat v,\hat S),
$$
where the Lagrangian $\hat \calL^\eps(\hat v,\hat S)$ is defined by
$$
\hat \calL^\eps(\hat v,\hat S):=\int_{\Ome}\hat S\cdot E\hat v- \frac 12(\hat\C^\eps)^{-1}\hat S\cdot\hat S-\hat F^\eps\cdot E\hat v-\hat b^\eps\cdot \hat v\,d\hat x-\langle \hat f^\eps,\hat v\rangle_{H^{1/2}_{00}(\pNOme)}.
$$
Since $\hat \calL_\eps$ satisfies the assumptions of the min-max Theorem (see e.g.\ \cite{ET1999} p.\ 176 Proposition 2.4 and Remark 2.4), it follows that
\begin{align*}
\inf_{\hat v\in \hat \calA^\eps}&\max_{\hat S\in L^2(\Ome;\R^{3\times3}_{\rm sym})}\hat \calL^\eps(\hat v,\hat S)=\max_{\hat S\in L^2(\Ome;\R^{3\times3}_{\rm sym})}\inf_{\hat v\in \hat \calA^\eps}\hat \calL^\eps(\hat v,\hat S)\\
&=\max_{\hat S\in L^2(\Ome;\R^{3\times3}_{\rm sym})}
\big(
\int_{\Ome}- \frac 12(\hat\C^\eps)^{-1}\hat S\cdot\hat S\,d\hat x\\
&\hspace{1,5cm}+
\inf_{\hat v\in \hat \calA^\eps}
\int_{\Ome}(\hat S-\hat F^\eps)\cdot E\hat v-\hat b^\eps\cdot \hat v\,d\hat x-\langle \hat f^\eps,\hat v\rangle_{H^{1/2}_{00}(\pNOme)}
\big),
\end{align*}
but, from \eqref{pgg}  we deduce that
\begin{align*}
\inf_{\hat v\in \hat \calA^\eps}&
\int_{\Ome}(\hat S-\hat F^\eps)\cdot E\hat v-\hat b^\eps\cdot \hat v\,d\hat x-\langle \hat f^\eps,\hat v\rangle_{H^{1/2}_{00}(\pNOme)}\\
&=\inf_{\hat v\in \hat \calA^\eps}\big(\int_{\Ome}(-\div(\hat S- \hat F^\eps)-\hat b^\eps)\cdot \hat v\,d\hat x 
+\langle (\hat S-\hat F^\eps) \hat n-\hat f^\eps,\hat v\rangle_{H^{1/2}_{00}(\pNOme)}\big)\\
&=\left\{
\begin{array}{ll}
0 & \mbox{if } \hat S\in \hat\calS^\eps,\\
-\infty & \mbox{otherwise},
\end{array}
\right.
\end{align*}
where we denote by
\begin{align*}
\hat \calS^\eps:=\{\hat S\in L^2(\Ome;\R^{3\times3}_{\rm sym}):\,\,& \div (\hat S- \hat F^\eps)+\hat b^\eps=0 \mbox{ in }L^2(\Ome;\R^3) \mbox{ and }\\
& (\hat S-\hat F^\eps) \hat n-\hat f^\eps=0 \mbox{ in }(H^{1/2}_{00}(\pNOme;\R^3))'\}
\end{align*}
the set of admissible stresses.

\color{black}

By defining the dual energy by
$$
\hat \calF^{\eps*}(\hat S):=\frac 12 \int_{\Ome}(\hat\C^\eps)^{-1} \hat S\cdot \hat S\,d\hat x,
$$
it  follows that
\begin{equation}\label{mineps}
\hat \calF^\eps(\hat u^\eps)=\inf_{\hat v\in \hat \calA^\eps}\hat \calF^\eps(\hat v)=-\min_{\hat S\in \hat \calS^\eps}\hat \calF^{\eps*}(\hat S)=:-\hat \calF^{\eps*}(\hat T^\eps),
\end{equation}
and that $\hat T^\eps=\hat \C^\eps E\hat u^\eps$. 

The minimization problem 
$$
\min_{\hat S\in \hat \calS^\eps}\hat \calF^{\eps*}(\hat S).
$$
is called \textsc{Dual Problem}.
\begin{remark}\label{dualityeps}
Note that the stress $\hat \sigma^\eps:=\hat \C^\eps E\hat w^\eps$ associated to
 the solution $\hat w^\eps$ of \eqref{pbleeps0}, see also \eqref{pbleeps}, is given by 
 $$\hat \sigma^\eps=\hat T^\eps+\hat \C^\eps E\hat {g}^\eps.$$
\end{remark}

\section{Rescaled problems}\label{sec4}

We now rescale the problems introduced in Section \ref{sec3} to a domain independent of $\eps$. 
To this end, we set 
$$
\Om:=\Om_1,\quad \pDOm:=\pNOm_1,\quad \pDOm:=\pNOm_1.
$$
We define the change of variables $p^\eps:\Om\to\Ome$ by 
$$
p^\eps(x_1,x_2,x_3):=(x_1,x_2,\eps x_3),
$$ 
and we let
$$
P^\eps:=\nabla p^\eps=\mbox{diag}\,(1,1,\eps).
$$ 
For $\hat v:\Ome\to \R^3$ we define $v:\Om\to\R^3$ by
\begin{equation}\label{vvhat}
v:=P^\eps \hat v\circ p^\eps,
\end{equation}
so that
$$
\nabla v=P^\eps (\nabla\hat v)\circ p^\eps P^\eps, \mbox{ and }E v=P^\eps (E\hat v)\circ p^\eps P^\eps.
$$
We denote by
\begin{equation}\label{10bis}
E^\eps v:=(P^\eps)^{-1}Ev(P^\eps)^{-1}=(E\hat v)\circ p^\eps.
\end{equation}

We assume that $\hat \C^\eps$, $\hat b^\eps$, $\hat H^\eps$, $\hat {g}^\eps$, and $\hat f^\eps$ are such that
\begin{equation}\label{defcb}
\hat\C^\eps\circ p^\eps=\C,\quad P^\eps \hat b^\eps\circ p^\eps=b,\quad \hat H^\eps\circ p^\eps=H, \quad  P^\eps \hat { g}^\eps\circ p^\eps=g,
\end{equation}
and that
\begin{equation}\label{idf}
\langle \hat f^\eps,\hat v\rangle_{H^{1/2}_{00}(\pNOme)}
=\eps \langle f, v\rangle_{H^{1/2}_{00}(\pNOm)}
\quad \mbox{ for every }\hat v\in H^{1/2}_{00}(\pNOme),
\end{equation}
for some coercive tensor field $\C \in L^\infty(\Om;\R^{3\times3\times3\times3}_{\rm sym})$, $b\in L^2(\Om;\R^3)$, $H\in  L^2(\Om;\R^{3\times3}_{\rm sym})$,
$ g\in H^{1}(\Om;\R^3)$ such that $(Eg)_{i3}=0$,  and $f\in(H^{1/2}_{00}(\pNOm;\R^3))'$.
From \eqref{39} we deduce that
\begin{equation}\label{39bis}
\hat F^\eps\circ p^\eps= H+\C E^{\eps}{g}=H+\C E{g} =:F.
\end{equation}

\begin{remark}\label{remgG}
The required condition  $(Eg)_{i3}=0$ is equivalent to say that $g$ is a Kirchhoff-Love displacement.
Indeed, this assumption and also those on $\hat \C^\eps, \hat B^\eps, \hat H^\eps,$ and $\hat f^\eps$  could be relaxed. For instance, it would be enough to require that
\begin{equation}\label{defcb2}
\hat H^\eps\circ p^\eps=\bar H^\eps, \quad  P^\eps \hat { g}^\eps\circ p^\eps=\bar g^\eps,
\end{equation}
for some  $\bar H^\eps\in  L^2(\Om;\R^{3\times3}_{\rm sym})$,
$\bar g^\eps\in H^{1}(\Om;\R^3)$
which further satisfy
\begin{equation}\label{convg}
\bar g^\eps \weak  \bar g\quad \mbox{in }L^2(\Om;\R^{3}),
\end{equation}
for some  $\bar g\in L^2(\Om;\R^{3})$,
and
\begin{equation}\label{convHG}
\bar H^\eps \to \bar H, \quad \C E^\eps\bar g^\eps\to\bar G\quad \mbox{in }L^2(\Om;\R^{3\times3}_{\rm sym}),
\end{equation}
for some $\bar H, \bar G\in L^2(\Om;\R^{3\times3}_{\rm sym})$.

We note though that from \eqref{convHG} we have that 
$$
E^\eps\bar g^\eps\to \C^{-1} \bar G\quad \mbox{in }L^2(\Om;\R^{3\times3}_{\rm sym}),
$$
which, combined with \eqref{convg}, Korn inequality and Rellich compactness Theorem,  implies that $(E \bar g)_{i3}=0$, i.e., that $\bar g$ is a Kirchhoff-Love displacement, and that convergence \eqref{convg} actually takes place in $H^1(\Om;\R^{3})$.
\end{remark}
\color{black}
\begin{remark}\label{remf2}
For the example $\hat {\bar f}^\eps$ considered in Remark \ref{remf}, with $\hat {\bar f}_\pm^\eps\in L^{2}(\om_\eps^\pm; \R^3)$ and $\hat {\bar f}_\ell^\eps\in L^2(\pNOme_\ell;\R^3)$ which satisfy 
$$
P^\eps \hat {\bar f}_\pm^\eps\circ p^\eps=:\eps {\bar f}^{\pm},\qquad \eps P^\eps \hat {\bar f}_\ell^\eps\circ p^\eps=:{\bar f}_\ell,
$$
for some ${\bar f}^{\pm}\in L^{2}(\om^\pm; \R^3)$ and ${\bar f}_\ell\in L^2(\pNOm^\ell;\R^3)$,
the rescaled surface load $\bar f$ defined by \eqref{idf} is given by 
\begin{align*}
\langle  {\bar f}, v\rangle_{H^{1/2}_{00}(\pNOm)}=& \, \int_{\om^+} {\bar f}_{+}\cdot v\,dx+\int_{\om^-} {\bar f}_{-}\cdot v\,dx+ \int_{\pNOm^\ell} {\bar f}_{\ell}\cdot v\,dx.
\end{align*}
\end{remark}

We define the \textsc{Rescaled Primal Problem} as
$$
\inf_{v\in \calA^\eps}\calF^\eps( v),
$$
where the set of rescaled admissible displacements and the rescaled energy are defined by
$$
\calA^\eps:=\{v\in H^1(\Om;\R^3):v = 0 \mbox{ on }\pDOm\},
$$
and
\begin{equation}\label{rpf}
\calF^\eps(v):=\int_{\Om}\frac 12\C E^\eps v\cdot E^\eps v\,-F\cdot E^\eps v
 +
b\cdot v\,dx-\langle f,v\rangle_{H^{1/2}_{00}(\pNOm)}.
\end{equation}
With the assumptions \eqref{defcb}-\eqref{39bis} we have
$$\hat \calF^\eps(\hat v)=\eps\calF^\eps( v),$$
where the relation between $v$ and $\hat v$ is given by \eqref{vvhat}.

\begin{remark}\label{GL}
In the line of Remark \ref{GLe}, we now make a comparison between the rescalings adopted for the ``generalized forces'' and the ``standard forces''. The rescaled ``generalized force'' $H$ contributes to the primal energy, see \eqref{rpf} and \eqref{39bis},  with the term 
\begin{equation}\label{HEe}
\int_\Om H\cdot E^\eps v\,dx,
\end{equation}
while the ``standard forces'' contribute with the terms
$$
\int_{\Om} b\cdot v\,d x+
\langle f, v\rangle_{H^{1/2}_{00}(\pNOm)}.
$$
In order to make a comparison we need to rewrite the contribution of the ``standard forces''
in a form similar to \eqref{HEe}. As in Remark \ref{GLe}, 
given $f\in(H^{1/2}_{00}(\pNOm;\R^3))'$ and $b\in L^2(\Ome;\R^3)$,
we may find $\bar H \in \Hdiv$ such that
 \begin{equation}\label{ggFl}
\int_{\Om} b\cdot v\,d x+
\langle f, v\rangle_{H^{1/2}_{00}(\pNOm)}=\int_{\Om} \bar H\cdot E v\,d x,
\end{equation}
for all $ v\in  \calA^\eps$. The right hand side of \eqref{ggFl} may be rewritten as
$$
\int_{\Om} \bar H\cdot E v\,d x=\int_{\Om} P^\eps \bar H P^\eps \cdot (P^\eps)^{-1}E v(P^\eps)^{-1}\,d x=\int_{\Om} P^\eps \bar H P^\eps \cdot E^\eps v\,d x,
$$
and the last term is exactly in the form of \eqref{HEe}. 
Since $(P^\eps \bar H P^\eps)_{i3}\to0$ in $L^2(\Om;\R^{3\times3}_{\rm sym})$, while, in general, $H_{i3}\ne0$ we deduce that the scaling of the ``standard forces'' is weaker than that applied to the ``generalized forces''.
\end{remark}

We now change variables to the dual problem.
Setting
$$
S:=\hat S\circ p^\eps,
$$
for any $\hat S\in\hat\calS^\eps$ and $\hat v\in \calA^\eps$ we have, from \eqref{pgg} and \eqref{10bis}, that
on one hand
\begin{align}
\int_{\Ome} (\hat S-\hat F^\eps)\cdot E \hat v\,d\hat x&=\eps\int_{\Om} (\hat S\circ p^\eps-F)
\cdot E^\eps  v\,d x\nonumber\\
&=\eps\int_{\Om} (P^\eps)^{-1}( S-F)(P^\eps)^{-1}\cdot E  v\,d x,\label{19bis}\\
&=-\eps\int_\Om\div((P^\eps)^{-1}(S-F)(P^\eps)^{-1})\cdot v\,dx\nonumber\\
&\hspace{1cm}+\eps\langle(P^\eps)^{-1}(S-F)(P^\eps)^{-1})n,v\rangle_{H^{1/2}_{00}(\pNOm)},\nonumber
\end{align}
while on the other hand
\begin{align}
\int_{\Ome} (\hat S-\hat F^\eps)\cdot E \hat v\,d\hat x&=
-\int_{\Ome}\div(\hat S-\hat F^\eps)\cdot \hat v\,d\hat x
+\langle(\hat S-\hat F^\eps)\hat n,\hat v\rangle_{H^{1/2}_{00}(\pNOme)} \nonumber\\
&=\int_{\Ome}\hat b^\eps\cdot \hat v\,d\hat x
+\langle\hat f^\eps,\hat v\rangle_{H^{1/2}_{00}(\pNOme)} \label{19tris}
\\
&=\eps\int_{\Om} b\cdot  v\,d x
+\eps\langle f, v\rangle_{H^{1/2}_{00}(\pNOm)}.\nonumber
\end{align}
Thus from the previous two equations we find that $S\in \calS^\eps$ if and only if
$$
\left\{
\begin{array}{ll}
 \div ((P^\eps)^{-1}( S-F)(P^\eps)^{-1})+b=0 &\mbox{ in }L^2(\Om;\R^3),\\
((P^\eps)^{-1}( S-F)(P^\eps)^{-1}) n= f &\mbox{ in }(H^{1/2}_{00}(\pNOm;\R^3))'.
\end{array}
\right.
$$

Hence, after rescaling the admissible set $\hat\calS^\eps$ becomes
\begin{align*}
\calS^\eps:=\{S\in L^2(\Om;\R^{3\times3}_{\rm sym}):&\, \div ((P^\eps)^{-1}( S-F)(P^\eps)^{-1})+b=0 \mbox{ in }L^2(\Om;\R^3)\\
& \mbox{ and } ((P^\eps)^{-1}( S-F)(P^\eps)^{-1}) n= f \mbox{ in }(H^{1/2}_{00}(\pNOm;\R^3))'\},
\end{align*}
the dual energy rewrites as
$$
\calF^{\eps*}(S):=\frac 12 \int_{\Om}\C^{-1} S\cdot  S\,d x,
$$
and the \textsc{Rescaled Dual Problem} is
$$
\inf_{ S\in  \calS^\eps} \calF^{\eps*}( S).
$$
\begin{remark}\label{relFstar}
With the notation above we have
$$
 \hat \calF^{\eps*}( \hat S)=\eps \calF^{\eps*}( S).
$$
In particular, it follows that if $T^\eps$ is the minimizer of $\calF^{\eps*}$, i.e.,
$$
\calF^{\eps*}( T^\eps)=\inf_{ S\in  \calS^\eps} \calF^{\eps*}( S),
$$
and if $\hat T^\eps$ is the minimizer of $ \hat \calF^{\eps*}$, see \eqref{mineps},
then
$$
T^\eps=\hat T^\eps\circ p^\eps.
$$

Let $w^\eps:=P^\eps\hat w^\eps\circ p^\eps$ be the rescaled displacement of the solution $\hat w^\eps$ of \eqref{pbleeps0}.
Then the rescaled stress $\sigma^\eps:=\hat \sigma^\eps\circ p^\eps=\C E^\eps w^\eps$ associated to
 the solution of \eqref{pbleeps0}, see Remark \ref{dualityeps}, is given by 
 $$\sigma^\eps= T^\eps+ \C E^\eps{g}=T^\eps+ \C E{g}.$$
\end{remark}

\begin{remark}
The rescaled dual problem coincides with the dual of the rescaled direct problem.
\end{remark}

\section{Gamma-convergence of the Rescaled Dual Functional}\label{sec5}

In this section, after studying the compactness of the dual problem in the weak-$L^2$ topology, we
identify the $\Gamma$-limit of the sequence of dual functionals. Moreover, we prove the strong convergence in the $L^2$ topology of the minimizers. For what follows it is useful to notice, see \eqref{19bis} and \eqref{19tris}, that 
$S\in\calS^\eps$ if and only if
\begin{equation}\label{blimitS}
\int_{\Om} ( S-F)\cdot E^\eps  v\,d x=\int_{\Om} b\cdot v\,d x+\langle f, v\rangle_{H^{1/2}_{00}(\pNOm)} ,
\end{equation}
for any $v\in \calA^\eps$.
From \eqref{blimitS} it easily follows that the set $\calS^\eps$ is not empty, indeed it can be shown that for every $\eps>0$ there exist $S^\eps\in\calS^\eps$ such that $\sup_\eps\|S^\eps\|_{L^2(\Om)}<+\infty$.
This, then implies that $\sup_\eps\calF^*_\eps(S^\eps)<+\infty.$

Before stating the compactness result it is convenient to set 
$$
KL_{0}(\Om):=\{v\in H^1(\Om;\R^3): (Ev)_{i3}=0, \mbox{ for }i=1,2,3, \mbox{ and }v=0 \mbox{ on }\pDOm\},
$$
and
\begin{align}
\calS:=\{S\in L^2(\Om;\R^{3\times3}_{\rm sym}):& \, S_{i3}=F_{i3}, \mbox{ for }i=1,2,3, \mbox{ and }\label{defS}\\
&\int_{\Om} ( S-F)\cdot E  w\,d x=\int_{\Om} b\cdot w\,d x+\langle f, w\rangle_{H^{1/2}_{00}(\pNOm)}\nonumber
\\&\hspace{4cm}\mbox{ for every }w\in KL_{0}(\Om)\}.\nonumber
\end{align}

\begin{lemma}\label{compactness}
Let $S^\eps\in\calS^\eps$ be a sequence such that $\sup_\eps\calF^*_\eps(S^\eps)<+\infty.$
Then there exist a subsequence, not relabeled, and an $S\in  \calS$ such that
$$S^\eps\weak S\mbox{ in }L^2(\Om;\R^{3\times3}_{\rm sym}).$$
\end{lemma}
\proof Let $c>0$ be such that $\C^{-1}(x)T\cdot T\ge c|T|^2$ for a.e.\ $x\in\Om$ and for every symmetric matrix $T$.
Thus
\begin{align*}
+\infty>\calF^{\eps*} (S^\eps)&\ge \frac 12 c \|S^\eps\|^2_{L^2(\Om)},
\end{align*}
and hence  $\sup_\eps\| S^\eps\|_{L^2(\Om)}<+\infty$, which implies that
there exist a subsequence, not relabeled, and an $S\in  L^2(\Om;\R^{3\times3}_{\rm sym})$ such that
$$S^\eps\weak S\mbox{ in }L^2(\Om;\R^{3\times3}_{\rm sym}).$$
Let $w\in KL_{0}(\Om)$ and $\psi\in C_0^\infty(\Om;\R^3)$.
Define 
\begin{align*}
v^\eps_\alpha(x_1,x_2,x_3)&:=w_\alpha(x_1,x_2,x_3)+\eps\int_0^{x_3}\psi_\alpha(x_1,x_2,s)\,ds,\\
v^\eps_3(x_1,x_2,x_3)&:=w_3(x_1,x_2,x_3)+\eps^2\int_0^{x_3}\psi_3(x_1,x_2,s)\,ds.
\end{align*}
Then $v^\eps\in \calA^\eps$, $v^\eps\to w$ in $H^1(\Om;\R^3)$ and
$$
E^\eps v^\eps=(P^\eps)^{-1}Ev^\eps(P^\eps)^{-1}\to Ew+
\left(\begin{array}{ccc}
0 & 0 &  \psi_1/2\\
& 0 & \psi_2/2\\
{\rm sym} & & \psi_3
\end{array}\right)
\mbox{ in } L^2(\Om;\R^{3\times3}).
$$
By taking $S=S^\eps$ and $v=v^\eps$ in \eqref{blimitS} and by passing to the limit, we find 
$$
\int_{\Om} ( S-F)\cdot Ew+ (S-F)e_3\cdot\psi\,d x=\int_{\Om} b\cdot w\,d x+\langle f, w\rangle_{H^{1/2}_{00}(\pNOm)}.
$$
Since $w$ and $\psi$ are arbitrary functions, in the respective domains, we easily conclude that $S\in\calS$.
\QED

We now identify the $\Gamma$-limit of the dual functionals.

\begin{theorem}\label{gammaconv}
The extended functional $\calF_{\rm ext}^\eps:L^2(\Om;\R^{3\times3}_{\rm sym})\to\R\cup\{+\infty\}$ defined by
$$
\calF_{\rm ext}^\eps(S)=\left\{
\begin{array}{cl}
\calF^{\eps*}(S) & \mbox{if }S\in \calS^\eps,\\
+\infty & \mbox{otherwise},
\end{array}
\right.
$$
sequentially $\Gamma$-converges with respect to the weak topology of $L^2(\Om;\R^{3\times3}_{\rm sym})$ to the functional
$$
\calF_{\rm ext}(S)=\left\{
\begin{array}{cl}
\calF^*(S) & \mbox{if }S\in\calS,\\
+\infty & \mbox{if }S\in L^2(\Om;\R^{3\times3}_{\rm sym})\setminus\calS,
\end{array}
\right.
$$
where 
$$
\calF^*(S):=\frac 12 \int_{\Om}\C^{-1} S\cdot  S\,d x.
$$
\end{theorem}
\proof
We need to prove that:
\begin{itemize}
\item[a)] for every $S\in L^2(\Om;\R^{3\times3}_{\rm sym})$ and every sequence $S^\eps\in L^2(\Om;\R^{3\times3}_{\rm sym})$ such that
$S^\eps\weak S$ in $L^2(\Om;\R^{3\times3}_{\rm sym})$ it holds
$$\liminf_\eps\calF_{\rm ext}^\eps(S^\eps)\ge \calF_{\rm ext}(S);$$
\item[b)] for every $S\in L^2(\Om;\R^{3\times3}_{\rm sym})$ there exists a sequence $S^\eps\in L^2(\Om;\R^{3\times3}_{\rm sym})$ such that $S^\eps\weak S$ in $L^2(\Om;\R^{3\times3}_{\rm sym})$ and
$$\limsup_\eps\calF_{\rm ext}^\eps(S^\eps)\le \calF_{\rm ext}(S).$$
\end{itemize}
We start by proving a). Let $S\in L^2(\Om;\R^{3\times3}_{\rm sym})$ and  $S^\eps\in L^2(\Om;\R^{3\times3}_{\rm sym})$ be a sequence such that $S^\eps\weak S$ in $L^2(\Om;\R^{3\times3}_{\rm sym})$. We may assume that 
$$\liminf_\eps\calF_{\rm ext}^\eps(S^\eps)=\lim_\eps\calF_{\rm ext}^\eps(S^\eps)<+\infty.$$
Then $\sup_\eps\calF_{\rm ext}^\eps(S^\eps)=\sup_\eps\calF^{\eps *}(S^\eps)<+\infty$ and hence, by Lemma \ref{compactness} it follows that $S\in \calS$. By a standard semicontinuity argument  we have
\begin{align*}
\liminf_\eps\calF_{\rm ext}^\eps(S^\eps)&=\lim_\eps\frac 12 \int_{\Om}\C^{-1} S^\eps\cdot  S^\eps\,d x\ge\frac 12 \int_{\Om}\C^{-1} S\cdot  S\,d x=\calF^{*}(S)=\calF_{\rm ext}(S).
\end{align*}

We now prove b), which is usually called the recovery sequence condition. Let $S\in L^2(\Om;\R^{3\times3}_{\rm sym})$. We may assume that $\calF_{\rm ext}(S)<+\infty$.
Thus $S\in \calS$. 
To construct the recovery we consider the following problem:
\begin{equation}\label{recvaru}
\left\{\begin{array}{l}
u^\eps\in \calA^\eps,\\
\displaystyle\int_{\Om}(\C E^\eps u^\eps+S-F)\cdot E^\eps \varphi\,dx=\int_{\Om} b\cdot \varphi\,d x+\langle f, \varphi\rangle_{H^{1/2}_{00}(\pNOm)},\mbox{ for every } \varphi\in \calA^\eps.
\end{array}
\right.
\end{equation}
By the definition of the operator $E^\eps$ and Korn's inequality we have that
$\|E^\eps \varphi\|_{L^2(\Omega)}\ge\|E \varphi\|_{L^2(\Omega)}\ge C \|\varphi\|_{H^1(\Omega)}$, for every $\varphi\in  \calA^\eps$ and for a constant $C$ independent of $\varphi$. This together with the positive definiteness of the elasticity tensor $\C$ implies that the solution $u^\eps$ of problem 
\eqref{recvaru} satisfies the bound:
\begin{equation}\label{B_Eue}
\sup_\eps\|E^\eps u^\eps\|_{L^2(\Omega)}<+\infty,
\end{equation}
and, as a consequence, $\sup_\eps\|u^\eps\|_{H^1(\Omega)}<+\infty$. Up to subsequences,
we have that
$$u^\eps\weak\check u \quad\mbox{ in }H^1(\Omega;\R^3),$$
for some $\check u\in H^1(\Omega;\R^3)$. By the definition of $E^\eps$, also
$$(E^\eps u^\eps)_{\alpha\beta}=(E u^\eps)_{\alpha\beta}\weak (E \check u)_{\alpha\beta} \quad\mbox{ in }L^2(\Omega), \quad \mbox{and}\quad (E u^\eps)_{i3}\to 0 \quad\mbox{ in }L^2(\Omega).$$
Whence $\check u\in KL_{0}(\Om)$. Moreover, up to a subsequence, we have that 
$$(E^\eps u^\eps)_{i3}\weak \check \psi_i \quad\mbox{ in }L^2(\Omega),$$
for some $\check \psi \in L^2(\Omega;\R^3)$.
These convergences can be compactly rewritten as
$$
E^\eps u^\eps \weak
\begin{pmatrix}
(E\check u)_{\alpha\beta} &\check \psi_\alpha\\
\check \psi_\beta & \eta_3
\end{pmatrix}
=:E(\check u, \check \psi) \quad\mbox{ in }L^2(\Omega;\R^{3\times 3}).
$$
Set
\begin{equation}\label{Seps}
S^\eps:=S+\C E^\eps u^\eps.
\end{equation}
That $S^\eps\in\calS^\eps$ follows from \eqref{blimitS} and \eqref{recvaru}, while, up to a subsequence,
\begin{equation}\label{diffS}
S^\eps\weak S+\C E(\check u, \check \psi)=:\check S \quad\mbox{in }L^2(\Om;\R^{3\times3}).
\end{equation}
Let $w\in  KL_{0}(\Om)$, $\eta \in C_0^\infty(\Om;\R^3)$, and set 
\begin{align*}
\varphi_\alpha(x_1,x_2,x_3)&:=w_\alpha(x_1,x_2,x_3)+\eps\int_0^{x_3}2\eta_\alpha(x_1,x_2,s)\,ds,\\
\varphi_3(x_1,x_2,x_3)&:=w_3(x_1,x_2,x_3)+\eps^2\int_0^{x_3}\eta_3(x_1,x_2,s)\,ds.
\end{align*}
Then,
$$
E^\eps\varphi=E(w,\eta)+R^\eps, \quad \mbox{with}\quad R^\eps\to 0 \mbox{ in }L^2(\Om;\R^{3\times3}),
$$
and with such a $\varphi$ we may pass to the limit in \eqref{recvaru} to find
\begin{equation}\label{recvaru1}
\int_{\Om}(\check S-F)\cdot E(w,\eta)\,dx=\int_{\Om} b\cdot w\,d x+\langle f, w\rangle_{H^{1/2}_{00}(\pNOm)},
\end{equation}
which holds for every $w\in  KL_{0}(\Om)$ and $\eta \in C_0^\infty(\Om;\R^3)$.

Since $S\in \calS$ we have, from the definition \eqref{defS} of $\calS$, that
\begin{equation}\label{recvaru2}
\int_{\Om}(S-F)\cdot E(w,\eta)\,dx=\int_{\Om} b\cdot w\,d x+\langle f, w\rangle_{H^{1/2}_{00}(\pNOm)},
\end{equation}
holds for every $w\in  KL_{0}(\Om)$ and $\eta \in C_0^\infty(\Om;\R^3)$. The difference between \eqref{recvaru1} and 
\eqref{recvaru2} delivers:
\begin{equation}\label{recvaru3}
\int_{\Om}(\check S-S)\cdot E(w,\eta)\,dx=0,
\end{equation}
for every $w\in  KL_{0}(\Om)$ and $\eta \in C_0^\infty(\Om;\R^3)$. By density this equation holds also
for every $\eta \in L^2(\Om;\R^3)$. Taking $w=\check u$, $\eta=\check \psi$, and using \eqref{diffS}
we obtain
$$
\int_{\Om}\C E(\check u, \check \psi)\cdot E(\check u, \check \psi)\,dx=0,
$$
which implies that $E(\check u, \check \psi)=0$ almost everywhere in $\Om$, and  consequently $\check \psi=0$,
and also $\check u=0,$  since $\check u\in KL_{0}(\Om)$. 
Now, taking $\varphi=u^\eps$ in \eqref{recvaru} and passing to the limit we deduce that
\begin{align*}
\lim_{\eps\to 0}\int_{\Om}\C E^\eps u^\eps\cdot E^\eps u^\eps\,dx
&=\lim_{\eps\to 0}\int_{\Om} -(S-F)\cdot E^\eps u^\eps +b\cdot u^\eps\,d x+\langle f, u^\eps\rangle_{H^{1/2}_{00}(\pNOm)}\\
&=\int_{\Om} -(S-F)\cdot E(\check u, \check \psi) +b\cdot \check u\,d x+\langle f, \check u\rangle_{H^{1/2}_{00}(\pNOm)}\\
&=0,
\end{align*}
thence $E^\eps u^\eps\to 0$ in $L^2(\Om;\R^{3\times3})$ and, by \eqref{Seps},
$$
S^\eps\to S \quad\mbox{in }L^2(\Om;\R^{3\times3}).
$$
Since $S^\eps\in\calS^\eps$  we find:
$$
\lim_{\eps\to0}\calF^{\eps*}(S^\eps)=\lim_{\eps\to0}\frac 12\int_{\Om}\C^{-1}S^\eps\cdot S^\eps\,dx=
\frac 12\int_{\Om}\C^{-1}S\cdot S\,dx=\calF^*(S)=\calF_{\rm ext}(S),
$$
and the proof is completed.
\QED

\begin{remark}
We remark that in the second part of the proof of Theorem \ref{gammaconv} we have indeed shown that: for every $S\in L^2(\Om;\R^{3\times3}_{\rm sym})$ there exists a sequence $S^\eps\in L^2(\Om;\R^{3\times3}_{\rm sym})$ such that $S^\eps\to S$ in $L^2(\Om;\R^{3\times3})$ and
$$\lim_{\eps\to0}\calF_{\rm ext}^\eps(S^\eps)=\calF_{\rm ext}(S).$$
\end{remark}

\begin{remark}
In our setting, by Proposition 8.10 of \cite{DM1993}, sequential $\Gamma$-convergence is equivalent to $\Gamma$-convergence.  
\end{remark}

In the next theorem we prove the strong convergence of the minimizers.

\begin{theorem}\label{strongconv}
Let $T^\eps$ be the minimizer of $\calF^{\eps*}$ and $T$ be the minimizer of $\calF^*$. Then
$$T^\eps\to T \mbox { in } L^2(\Om;\R^{3\times3}_{\rm sym}),$$
and
$$\lim_{\eps\to0}\calF^{\eps*}(T^\eps)=\calF^*(T).$$
\end{theorem}

\proof
Let $T^\eps$ be the minimizer of $\calF^{\eps*}$. Then by Lemma \ref{compactness} we have that, up to a subsequence, 
$T^\eps\weak T$ in $L^2(\Om;\R^{3\times3}_{\rm sym})$, for some $T\in \calS$. Let $S\in \calS$ and let $S^\eps\in\calS^\eps$ be a sequence such that $\limsup_{\eps\to0} \calF^{\eps*}(S^\eps)\le \calF^*(S)$, which exists by Theorem \ref{gammaconv}.
Since $\calF^{\eps*}(T^\eps)\le\calF^{\eps*}(S^\eps)$, by Theorem \ref{gammaconv} we have
$$
\calF^*(T)\le \liminf_{\eps\to0} \calF^{\eps*}(T^\eps)\le \limsup_{\eps\to0} \calF^{\eps*}(T^\eps)\le \limsup_{\eps\to0} \calF^{\eps*}(S^\eps)\le \calF^*(S),$$
 which implies that $T$ is a minimizer of $\calF$, and by taking $S$ equal to $T$, that
$$\lim_{\eps\to0}\calF^{\eps*}(T^\eps)=\calF^*(T).$$
Since $\calF^*$ has a unique minimizer we have that the full sequence $T^\eps$ weakly converges to $T$ in $L^2(\Om;\R^{3\times3}_{\rm sym})$.
By convexity it then follows that $T^\eps\to T$ in $L^2(\Om;\R^{3\times3}_{\rm sym})$.
Indeed, we have
\begin{align*}
\lim_{\eps\to0}\int_{\Om}\C^{-1}(T^\eps-T)\cdot & (T^\eps-T)\,dx\\
&=2\lim_{\eps\to0}\big(
\calF^{\eps*}(T^\eps)-
\int_{\Om}\C^{-1}T^\eps\cdot T\,dx+\calF^*(T)\big)
=0,
\end{align*}
from which the strong convergence follows.
\QED

\begin{remark}\label{relFstar2}
The rescaled stress $\sigma^\eps=T^\eps+ \C E{g}$ associated to
 the solution of \eqref{pbleeps0}, see Remark \ref{relFstar}, strongly converges 
in $L^2(\Om;\R^{3\times3}_{\rm sym})$ to $\sigma:=T+ \C E{g}.$
\end{remark}

The next lemma,  similar to a result contained in \cite{BGK12}, allows us to 
characterize the minimizing stress tensor.
\begin{lemma}\label{orth}
Let $D$ be a bounded, open subset of $\R^3$ with Lipschitz boundary $\partial D$.
Let $\partial_DD\ne \emptyset$ be the union of a finite number of open connected sets of $\partial D$.
Let 
$$
KL_{0}(D):=\{v\in H^1(D;\R^3): (Ev)_{i3}=0, \mbox{ and }v=0 \mbox{ on }\partial_DD\},
$$
\begin{align*}
\calK:=\{E\in L^2(D;\R^{3\times3}_{\rm sym}):&\exists z\in KL_0(D) \mbox{ and }\psi\in L^2(D;\R^3)\mbox{ such that }\\
&E=\left(
\begin{array}{cc}
(Ez)_{\alpha\beta} & \psi_{\beta}\\
\psi_{\alpha} &\psi_{3}
\end{array}
\right)
\},
\end{align*}
and
\begin{align*}
\calM=\{S\in L^2(D;\R^{3\times3}_{\rm sym}):& \, S_{i3}=0, \mbox{ and }\label{defSF0}\\
&\int_{D} S\cdot E  z\,d x=0 \mbox{ for every }z\in KL_{0}(D)\}.\nonumber
\end{align*}
Then
\begin{align*}
\calK=\calM^\perp.
\end{align*}

\end{lemma}
\proof 
We first note that $\calK$ is a closed subset of $L^2(D;\R^{3\times3}_{\rm sym})$.
Indeed, let $\{E^j\}\subset\calK$ be such that $E^j \to E$ in $L^2(D;\R^{3\times3}_{\rm sym})$. Then there exist $z^j\in KL_0(D)$ and $\psi^j\in L^2(D;\R^3)$ such that 
$$(Ez^j)_{\alpha\beta} \to (E)_{\alpha\beta}\qquad \psi^j_{i}\to\psi_{i}=(E)_{i3},\quad \mbox{ in }L^2(D),$$
for some  $\psi_i\in L^2(D)$.
Thus to show that $\calK$ is closed it suffices to show that there exists a $z\in KL_0(D)$
such that $(E)_{\alpha\beta}=(Ez)_{\alpha\beta}$. But since $z^j\in KL_0(D)$
we have that $Ez^j$ is a Cauchy sequence in $L^2(D;\R^{3\times3}_{\rm sym})$ and hence,
from Korn's inequality we deduce, in the components of 
$D$ whose boundary contain part of $\partial_DD$,  that $z^j\to z$ in the $H^1$ norm, while on the other components it
is $z^j$ minus its orthogonal projection on the set of infinitesimal rigid displacements which converges to some $z$ in the $H^1$ norm. Throughout $D$ we then have  $(E)_{\alpha\beta}=(Ez)_{\alpha\beta}$.

The proof of the lemma now follows easily. In fact, we have $\calK\subset \calM^\perp$ and $\calK^\perp\subset \calM$.
This latter inclusion implies that $\calM^\perp\subset (\calK^\perp)^\perp$. 
Hence
$$\calK\subset\calM^\perp\subset (\calK^\perp)^\perp,$$
but since $\calK$ is a closed subset of $L^2(D;\R^{3\times3}_{\rm sym})$ we have that
$(\calK^\perp)^\perp=\calK$.
\QED

\begin{theorem}
The minimizer  $T$ of $\calF^*$ satisfies the following problem:
\begin{equation}\label{ELT}
\left\{\begin{array}{l}
T\in \calS,\\
\displaystyle\int_{\Om}\C^{-1}T\cdot \Sigma\,dx=0,\quad \mbox{for every } \Sigma\in\calS_{0},
\end{array}
\right.
\end{equation}
where 
\begin{align}
\calS_{0}:=\{S\in L^2(\Om;\R^{3\times3}_{\rm sym}):& \, S_{i3}=0, \mbox{ and }\label{defSF0}\\
&\int_{\Om} S\cdot E  z\,d x=0 \mbox{ for every }z\in KL_{0}(\Om)\}.\nonumber
\end{align}
Moreover, there exist a unique $\psi\in L^2(\Om;\R^3)$  and a unique $u\in KL_{0}(\Om)$
such that
\begin{equation}\label{structureT}
T=\C\left(
\begin{array}{cc}
(Eu)_{\alpha\beta} & \psi_{\beta}\\
\psi_{\alpha} &\psi_{3}
\end{array}
\right).
\end{equation}
\end{theorem}

\proof
Problem \eqref{ELT} is simply the Euler-Lagrange equation of the problem $\inf_{S\in\calS}\calF^*(S)$.
From  \eqref{ELT} we have that
$$\C^{-1}T\in (\calS_{0})^\perp,$$
and hence from Lemma \ref{orth} we deduce that there exist $u\in KL_0(\Om)$ and $\psi\in L^2(\Om;\R^3)$ such that 
$$\C^{-1}T=\left(
\begin{array}{cc}
(Eu)_{\alpha\beta} & \psi_{\beta}\\
\psi_{\alpha} &\psi_{3}
\end{array}
\right).
$$
\QED
\begin{remark}\label{relFstar3}
The stress $\sigma=T+ \C E{g}$, limit of the stresses associated  to
 the solutions of \eqref{pbleeps0}, see Remark \ref{relFstar2}, 
 is given by
 $$\sigma=\C\left(
\begin{array}{cc}
(Eu+E\bar g)_{\alpha\beta} & \psi_{\beta}\\
\psi_{\alpha} &\psi_{3}
\end{array}
\right).
$$
Setting
$$w:=u+g\in KL_{g}(\Om):=\{v\in H^1(\Om;\R^3): (Ev)_{i3}=0,  \mbox{ and }v=g \mbox{ on }\pDOm\},
$$
we may write
 $$\sigma=\C\left(
\begin{array}{cc}
(Ew)_{\alpha\beta} & \psi_{\beta}\\
\psi_{\alpha} &\psi_{3}
\end{array}
\right).
$$

The rescaled stresses $\sigma^\eps=\C E^\eps w^\eps$
 strongly converge
in $L^2(\Om;\R^{3\times3}_{\rm sym})$ to $\sigma,$ see Remarks \ref{relFstar} and \ref{relFstar2},
thus
$$
E^\eps w^\eps\to
\left(
\begin{array}{cc}
(Ew)_{\alpha\beta} & \psi_{\beta}\\
\psi_{\alpha} &\psi_{3}
\end{array}
\right),\quad \mbox{ in }L^2(\Om;\R^{3\times3}).
$$
\end{remark}

\section{The bi-dimensional limit problem}\label{sec6}

The limit problem obtained in Section \ref{sec5} is defined on a three-dimensional domain. The aim of this Section is to show that it can be rewritten on a two-dimensional domain.

For a given $S\in\calS$ let
$$
S^N:=\int_{-1/2}^{1/2}S_{\alpha\beta}\,dx_3 e_\alpha\otimes e_\beta, 
\mbox{ and }
S^M:=\int_{-1/2}^{1/2}x_3 S_{\alpha\beta}\,dx_3 e_\alpha\otimes e_\beta.
$$
Similarly, using the components $F_{\alpha\beta}$, we define $F^N$ and $F^M$.
Let
$$
H^1_{0,D}(\omega;\R^2):=\{\eta\in H^1(\omega;\R^2):\eta=0 \mbox{ on }\pDom\},
$$
and
$$
H^2_{0,D}(\omega):=\{\eta\in H^2(\omega):\eta=\partial_n\eta=0 \mbox{ on }\pDom\}.
$$
For every $z\in KL_0(\Om)$ there exist  $(\eta_1,\eta_2)\in H^1_{0,D}(\omega;\R^2)$, $\eta_3\in H^2_{0,D}(\omega)$ such that 
$$
\left\{\begin{array}{l}
z_\alpha(x_1,x_2,x_3)=\eta_\alpha(x_1,x_2)-x_3\partial_\alpha \eta_3(x_1,x_2),\\
z_3(x_1,x_2,x_3)=\eta_3(x_1,x_2).
\end{array}
\right.
$$
A simple calculation shows that 
$$Ez=((E\eta)_{\alpha\beta}-x_3 \partial_\alpha\partial_\beta\eta_3)e_\alpha\otimes e_\beta,$$
and hence the condition, which also appears in the definition of $\calS$, see \eqref{defS},
$$
\int_{\Om} ( S-F)\cdot E  z\,d x=\int_{\Om} b\cdot z\,d x+\langle f, z\rangle_{H^{1/2}_{00}(\pNOm)},
$$
for every $z\in KL_{0}(\Om)$, rewrites as
$$
\int_{\omega} ( S^N-F^N)_{\alpha\beta}(E \eta)_{\alpha\beta}- 
( S^M-F^M)_{\alpha\beta}\partial_\alpha \partial_\beta\eta_3 \,d x=\calW^N((\eta_1,\eta_2))+\calW^M(\eta_3),
$$
where
$$
\calW^N((\eta_1,\eta_2)):=\int_\om \int_{-1/2}^{1/2}b_\alpha\,dx_3\,\eta_\alpha\,dx+\langle f_\alpha, \eta_\alpha\rangle_{H^{1/2}_{00}(\pNOm)},
$$
and
\begin{align*}
\calW^M(\eta_3):=& \int_\om \int_{-1/2}^{1/2}b_3\,dx_3\,\eta_3\,dx+\langle f_3, \eta_3\rangle_{H^{1/2}_{00}(\pNOm)}\\
&+\int_\om \int_{-1/2}^{1/2}x_3 b_\alpha\,dx_3\,\partial_\alpha\eta_3\,dx+\langle f_\alpha,x_3 \partial_\alpha\eta_3\rangle_{H^{1/2}_{00}(\pNOm)}.
\end{align*}
\begin{remark}\label{remf3}
If $\bar f$ is as in Remark \ref{remf2}, then the work done by the loads can be written more explicitly, for instance
$$
\langle \bar f_\alpha, \eta_\alpha\rangle_{H^{1/2}_{00}(\pNOm)}=
 \int_{\om^+} \bar f_{+\alpha} \,\eta_\alpha\,dx+\int_{\om^-} \bar f_{-\alpha}\, \eta_\alpha\,dx+ \int_{\pNom}
 \int_{-1/2}^{1/2} \bar f_{\ell\alpha}\,dx_3\, \eta_\alpha\,dx.
$$
\end{remark}

We therefore have
\begin{align}
\calS:=\{S\in L^2(\Om&;\R^{3\times3}_{\rm sym}): \, S_{i3}=F_{i3}, \mbox{ for }i=1,2,3, \mbox{ and }\label{defS2}\\
&\int_{\omega} ( S^N-F^N)\cdot E  \varphi\,d x=\calW^N(\varphi)\mbox{ for every }\varphi\in H^1_{0,D}(\omega;\R^2),\nonumber\\
&\int_{\omega} ( S^M-F^M)\cdot \nabla\nabla \psi\,d x=\calW^M(\psi) \mbox{ for every }\psi\in H^2_{0,D}(\omega).\nonumber
\}
\end{align}

We now rewrite the functional $\calF^*$ in terms of $S^N$ and $S^M$. To do so we let
\begin{align*}
\calL:=\{S\in L^2(\Om;\R^{3\times3}_{\rm sym}):& \,
\exists A, B\in L^2(\omega;\R^{2\times2}_{\rm sym})  \mbox{ such that}  \\
&
S_{\alpha\beta}(x_1,x_2,x_3)=A_{\alpha\beta}(x_1,x_2)+x_3B_{\alpha\beta}(x_1,x_2)
\}.
\end{align*}
Since $\calL$ is a closed subspace of $L^2(\Om;\R^{3\times3}_{\rm sym})$ we have
$$
L^2(\Om;\R^{3\times3}_{\rm sym})=\calL\oplus\calL^\perp.
$$
We note that $\Sigma\in\calL^\perp$ if and only if $\Sigma^N=\Sigma^M=\Sigma_{i3}=0$. Let $\Pi$ be the projection of $L^2(\Om;\R^{3\times3}_{\rm sym})$ onto $\calL$.  Then, from the relation
$$
\int_\Omega \Pi(S)\cdot\Sigma\,dx= \int_\Omega S\cdot\Sigma\,dx\mbox{ for every }\Sigma\in \calL,
$$
we infer that
$$
 \Pi(S)_{\alpha\beta}=S^N_{\alpha\beta}+12x_3 S^M_{\alpha\beta},\qquad\Pi(S)_{i3}=S_{i3}.
$$
Hereafter we denote by
$$S^\calL:=\Pi(S)\quad \mbox{ and }\quad S^c:=S-S^\calL.$$
and by
$$\calS^\calL:=\Pi(\calS)\quad \mbox{ and }\quad \calS^c:=\calS-\calS^\calL.$$
\begin{lemma}\label{ScLp}
With the notation just introduced we have that
$$\calS^c=\calL^\perp.$$
\end{lemma}

\proof
From the definition of $\calS^c$ it immediately follows that $\calS^c\subset\calL^\perp$.
To prove the opposite inclusion first note that
\begin{equation}\label{inc1}
\calS^{\calL}\subset \calS.
\end{equation}
Indeed, let $S^\calL\in\calS^\calL$ . Then there exists $S\in \calS$ such that $S^\calL=\Pi(S)$, that is
$(S^\calL)_{i3}=S_{i3}=F_{i3}$, and since $\Pi(S)_{\alpha\beta}=S^N_{\alpha\beta}+12x_3 S^M_{\alpha\beta}$ we have also that $(S^\calL)^N=S^N$ and $(S^\calL)^M=S^M$. Hence \eqref{inc1} follows from the representation of $\calS$ given in \eqref{defS2}.

Let $\Sigma\in \calL^\perp$. Let $S^\calL$ be any element of $\calS^\calL$. The condition  $\Sigma\in \calL^\perp$ implies that
$\Sigma^N=\Sigma^M=\Sigma_{i3}=0$ and hence we have, using \eqref{inc1}, that $\Sigma+S^\calL\in \calS$.  Since $\Pi(\Sigma)=0$, $\Pi(S^\calL)=S^\calL$ and the linearity of $\Pi$, which holds because $\calL$ is a closed linear subspace, we have
\begin{align*}
\Sigma=\Sigma+S^\calL-(\Pi(\Sigma)+\Pi(S^\calL))=\Sigma+S^\calL-\Pi(\Sigma+S^\calL)\in \calS\setminus\Pi(\calS)=\calS^c,
\end{align*}
and hence $\calL^\perp\subset\calS^c$.
\QED

We may therefore write
\begin{align*}
\calF^*(S)&=\frac 12 \int_{\Om}\C^{-1} S^\calL \cdot S^\calL+2 \C^{-1} S^\calL\cdot  S^c+\C^{-1} S^c\cdot  S^c\,d x\\
&=\calF^\perp(S^\calL,S^c)+\frac 12 \int_{\Om}\C^{-1} S^\calL \cdot S^\calL\,d x,
\end{align*}
where we have set
$$
\calF^\perp(S^\calL,S^c):=
 \int_{\Om} \C^{-1} S^\calL\cdot  S^c+\frac 12 \C^{-1} S^c\cdot  S^c\,d x.
$$
Thus, thanks to Lemma \ref{ScLp}, we have that
\begin{align*}
\inf_{S\in\calS}\calF^*(S)=\inf_{S^\calL\in\calS^\calL}\inf_{S^c\in\calL^\perp}
\calF^\perp(S^\calL,S^c)+\frac 12 \int_{\Om}\C^{-1} S^\calL \cdot S^\calL\,d x,
\end{align*}
and setting
$$
f^\perp(S^\calL):=\inf_{S^c\in\calL^\perp}\calF^\perp(S^\calL,S^c),
$$
we have
$$
\inf_{S\in\calS}\calF^*(S)=\inf_{S^\calL\in\calS^\calL}\calF^*_\calL(S^\calL),
$$
where we have set
\begin{equation}\label{FstarL}
\calF^*_\calL(S^\calL):=\frac 12 \int_{\Om}\C^{-1} S^\calL \cdot S^\calL\,d x+f^\perp(S^\calL).
\end{equation}
It is possible, even for a generic elasticity tensor $\C$, to write the function $f^\perp$ explicitly, but, as it can be seen in the next Theorem,  the explicit form of $f^\perp$ is quite involved.
\begin{theorem}\label{charfperp}
Let 
\begin{equation}\label{A1}
\bsc_{ij}:=\C_{i3j3}, \qquad \bar \C_{\alpha\beta\gamma\delta}:=\C_{\alpha\beta\gamma\delta}-\C_{\alpha\beta j3}\bsc^{-1}_{ji}\C_{i3\gamma\delta},
\end{equation}
\begin{equation}\label{A2}
\bar\C^{(i)}:=\int_{-1/2}^{1/2}x_3^i\,\bar\C\,dx_3\quad\mbox{for }i=0,1,2,
\end{equation}
\begin{equation}\label{A3}
\hat \C:=12 (\bar\C^{(2)}-\bar\C^{(1)}(\bar\C^{(0)})^{-1}\bar\C^{(1)}),
\end{equation}
and
\begin{eqnarray}
&\C^{nn}:=(\bar\C^{(0)})^{-1}+12(\bar\C^{(0)})^{-1}\bar\C^{(1)}\hat \C^{-1}\bar\C^{(1)}  (\bar\C^{(0)})^{-1},\label{A4}\\
&\C^{nm}:=-12(\bar\C^{(0)})^{-1}\bar\C^{(1)}\hat \C^{-1},\label{A4b}\\
&\C^{mn}:=-\hat \C^{-1}\bar\C^{(1)}  (\bar\C^{(0)})^{-1},\quad \C^{mm}:=\hat \C^{-1}.\label{A4t}
\end{eqnarray}
For a given $S^\calL\in\calS^\calL$,  let $\Lambda\in \calL^\perp$ be the minimizer of $\calF^\perp(S^\calL,\cdot)$, i.e., 
$$
f^\perp(S^\calL)=\inf_{S^c\in\calL^\perp}\calF^\perp(S^\calL,S^c)=\calF^\perp(S^\calL,\Lambda).
$$
Then $\Lambda=\C Z-S^\calL$ where\footnote{$a\odot b:=\frac 12 (a\otimes b +b\otimes a)$} $Z=\bar Z +z\odot e_3$, with $\bar Z:= Z^N+12x_3 Z^M$,
\begin{equation}\label{A5}
\left\{
\begin{array}{l}
Z^N:=\C^{nn} (S^\calL)^N+\C^{nm} (S^\calL)^M+\bsz^n,\\
Z^M:=\C^{mn} (S^\calL)^N+\C^{mm} (S^\calL)^M+\bsz^m,\\
z_j:=\bsc^{-1}_{ji}(F_{i3}-(\C\bar Z)_{i3}),
\end{array}
\right.
\end{equation}
and where
\begin{equation}\label{A6}
\bsz^m:=\hat \C^{-1}(\bar\C^{(1)}  (\bar\C^{(0)})^{-1}\bsf^N-\bsf^M), \quad
\bsz^n:=-(\bar\C^{(0)})^{-1}(\bsf^N+12 \bar\C^{(1)}\bsz^m),
\end{equation}
with
\begin{equation}\label{A7}
\bsf_{\alpha\beta}:=\C_{\alpha\beta j3}\bsc^{-1}_{ji}F_{i3}.
\end{equation}
Moreover
\begin{equation}\label{A8}
f^\perp(S^\calL)=\frac 12 \int_{\Om} \bar \C \bar Z\cdot \bar Z-\C^{-1} S^\calL\cdot  S^\calL\,d x+c,
\end{equation}
where the constant $c$ depends only on $F_{i3}$ and $\C$.
\end{theorem}
The proof of Theorem \ref{charfperp} is given in the Appendix at the end of the paper.
\begin{remark}
We note that:
\begin{enumerate}
\item if $\C_{\alpha\beta\gamma3}=\C_{\alpha333}=0$, i.e., triclinic symmetry, then
$$
\bar \C_{\alpha\beta\gamma\delta}=\C_{\alpha\beta\gamma\delta}-\frac{\C_{\alpha\beta 33}\C_{33\gamma\delta}}{\C_{3333}};
$$
\item if the material is triclinic and $F_{\alpha3}=0$ then $\Lambda_{\alpha3}=0$, i.e., the shear stresses are equal to zero. Indeed we have $\Lambda_{\alpha3}=\C_{\alpha3jk}Z_{jk}-S^\calL_{\alpha3}=
2\C_{\alpha3\beta3}Z_{\beta 3}=2\C_{\alpha3\beta3}z_{\beta}$, but since $\bsc_{\beta3}=0$ it follows that $z_\beta=\bsc^{-1}_{\beta\alpha}(F_{\alpha3}-(\C\bar Z)_{\alpha3})=-\bsc^{-1}_{\beta\alpha}\C_{\alpha3\gamma\delta}\bar Z_{\gamma\delta}=0$;
\item\label{rm6.3.2} if $\C(x_1, x_2, \cdot)$ is even, for almost every $(x_1,x_2)\in \om$, then $\bar\C^{(1)}$ is null, and hence
$$
\left\{
\begin{array}{l}
Z^N:=(\bar\C^{(0)})^{-1} (S^\calL)^N-(\bar\C^{(0)})^{-1}\bsf^N,\\
Z^M:=\frac{1}{12}(\bar\C^{(2)})^{-1} (S^\calL)^M-\frac{1}{12}(\bar\C^{(2)})^{-1} \bsf^M;
\end{array}
\right.
$$
\item\label{rm6.3.3} if $F_{i3}=0$ then $\bsf$, $\bsz^n$ and $\bsz^m$ are null matrices;
\item if $\C$ is independent of $x_3$ and $F_{i3}=0$ then items \ref{rm6.3.2}. and \ref{rm6.3.3}. of the present Remark hold and moreover 
\begin{equation}\label{A8b}
f^\perp(S^\calL)=\frac 12 \int_{\Om} \bar \C^{-1} S^\calL\cdot  S^\calL-\C^{-1} S^\calL\cdot  S^\calL\,d x+c.
\end{equation}
In fact, under these assumptions, we find 
$\bar\C^{(0)}=\bar \C$, $\bar\C^{(2)}=\frac{1}{12}\bar \C$ and hence $Z^N=\bar\C^{-1} (S^\calL)^N$,
$Z^M=\bar\C^{-1} (S^\calL)^M$, from which it follows that
$$\bar Z=\bar\C^{-1}((S^\calL)^N+12x_3(S^\calL)^M)=\bar\C^{-1}S^\calL.$$
Thus from the equation of $f^\perp$ given in Theorem \ref{charfperp} it follows the representation of $f^\perp$ given in \eqref{A8b}.
The constant $c$, see Appendix, is equal to zero if $F_{i3}=0$. Thus under these assumptions we have that, see \eqref{FstarL},
$$
\calF^*_\calL(S^\calL):=\frac 12 \int_{\Om}\bar\C^{-1} S^\calL \cdot S^\calL\,d x.
$$
\end{enumerate}
\end{remark}

Let $T^\calL$ be the minimizer of $\calF^*_\calL$, i.e.,
$$
\calF^*_\calL(T^\calL)=\inf_{S^\calL\in\calS^\calL}\calF^*_\calL(S^\calL),
$$
and $T^c\in \calS^c$ be the minimizer of $\calF^\perp(T^\calL,\cdot)$, i.e.,
$$
\calF^\perp(T^\calL,T^c)=\inf_{S^c\in\calL^\perp}\calF^\perp(T^\calL,S^c),
$$
then the minimizer of  $\calF^*$ is
$$
T=T^\calL+T^c.
$$
We note that once $T^\calL$ is known one can determine $T^c$ directly from Theorem \ref{charfperp}.

We conclude the section by noticing that the functional $\calF^*_\calL$, despite its appearance,
is essentially defined on $\om$.
\section{Appendix}
This appendix is devoted to the proof of Theorem \ref{charfperp}. Let $S^\calL$ be given and let $\Lambda\in \calL^\perp$ be the minimizer of $\calF^\perp(S^\calL,\cdot)$, i.e., 
$$
f^\perp(S^\calL)=\inf_{S^c\in\calL^\perp}\calF^\perp(S^\calL,S^c)=\calF^\perp(S^\calL,\Lambda).
$$
Then $\Lambda$ satisfies the following problem:
$$
\int_\Om \C^{-1}(S^\calL+\Lambda)\cdot \Sigma\,dx=0, \quad\mbox{for every }\Sigma\in \calL^\perp,
$$
that is
$$
Z:=\C^{-1}(S^\calL+\Lambda)\in \calL.
$$
Hence $\C Z=S^\calL+\Lambda$ and since $\Lambda\in \calL^\perp$ we have that
\begin{equation}\label{systemZ}
\left\{
\begin{array}{l}
(\C Z)_{i3}=(S^\calL)_{i3}=F_{i3},\\
(\C Z)^N=(S^\calL)^N,\\
(\C Z)^M=(S^\calL)^M.
\end{array}
\right.
\end{equation}
We now show that system \eqref{systemZ} delivers $Z$ uniquely. 
Let 
$$z_\alpha:=2Z_{\alpha3}, \quad z_3:=Z_{33},\quad \bar Z=Z_{\alpha\beta}e_\alpha\otimes e_\beta,$$
then we have
$$Z=\bar Z+z\odot e_3.$$
The first equation of \eqref{systemZ} rewrites as
$$(\C\bar Z)_{i3}+(\C z\odot e_3)_{i3}=F_{i3}$$
and by denoting, see \eqref{A1},
$$\bsc_{ij}:=\C_{i3j3},$$
it can be rewritten as
$$
(\bsc z)_i=F_{i3}-(\C\bar Z)_{i3}.
$$
Since $\C$ is positive definite we have that $\bsc$ is also positive definite, and hence
\begin{equation}\label{zj}
z_j=\bsc^{-1}_{ji}(F_{i3}-(\C\bar Z)_{i3}).
\end{equation}
We now evaluate the in-plane components of $\C Z$. We have
\begin{align*}
(\C Z)_{\alpha\beta}&=\C_{\alpha\beta\gamma\delta}\bar Z_{\gamma\delta}+\C_{\alpha\beta j3}z_j
=\C_{\alpha\beta\gamma\delta}\bar Z_{\gamma\delta}+\C_{\alpha\beta j3}\bsc^{-1}_{ji}(F_{i3}-(\C\bar Z)_{i3})\\
&= (\C_{\alpha\beta\gamma\delta}-\C_{\alpha\beta j3}\bsc^{-1}_{ji}\C_{i3\gamma\delta})\bar Z_{\gamma\delta}+\C_{\alpha\beta j3}\bsc^{-1}_{ji}F_{i3}.
\end{align*}
Setting, see \eqref{A1} and \eqref{A7},
$$\bar \C_{\alpha\beta\gamma\delta}:=\C_{\alpha\beta\gamma\delta}-\C_{\alpha\beta j3}\bsc^{-1}_{ji}\C_{i3\gamma\delta}, \qquad \bsf_{\alpha\beta}:=\C_{\alpha\beta j3}\bsc^{-1}_{ji}F_{i3},$$
we have
\begin{equation}\label{CZ}
(\C Z)_{\alpha\beta}=(\bar\C \bar Z)_{\alpha\beta}+\bsf_{\alpha\beta}.
\end{equation}
But, since $Z\in\calL$, we can write
$$\bar Z=Z^N+12 x_3 Z^M,$$
and, with this position, the second and third equations of \eqref{systemZ} rewrite as:
\begin{equation}\label{systemZ2}
\left\{
\begin{array}{l}
\bar\C^{(0)} Z^N+12\bar\C^{(1)} Z^M=(S^\calL)^N-\bsf^N,\\
\bar\C^{(1)} Z^N+12\bar\C^{(2)} Z^M=(S^\calL)^M-\bsf^M,
\end{array}
\right.
\end{equation}
where we have set, see \eqref{A2},
$$
\bar\C^{(i)}:=\int_{-1/2}^{1/2}x_3^i\,\bar\C\,dx_3\quad\mbox{for }i=0,1,2.
$$
Thanks to Lemma \ref{positive}, below, we have
\begin{equation}\label{ZN1}
Z^N=-12 (\bar\C^{(0)})^{-1}\bar\C^{(1)} Z^M+  (\bar\C^{(0)})^{-1}((S^\calL)^N-\bsf^N),
\end{equation}
and
\begin{equation}\label{ZM1}
Z^M=\hat \C^{-1}((S^\calL)^M-\bsf^M -\bar\C^{(1)}  (\bar\C^{(0)})^{-1}((S^\calL)^N-\bsf^N)),
\end{equation}
where $\hat \C$ is defined by \eqref{A3}.
\begin{lemma}\label{positive}
Let $c_\C>0$ be a constant such that
 $$
 \mbox{\rm essinf}_{\,x\in \Om}\, \C(x) A\cdot A\ge c_\C |A|^2,
 $$
 for every symmetric matrix $A\in \R^{3\times 3}$.

With the notation introduced above we have
$$
\bar \C \bar A\cdot \bar A=\min_{b\in\R^3}\C(\bar A+b\odot e_3)\cdot (\bar A+b\odot e_3)\ge c_\C |\bar A|^2
$$
for every symmetric matrix $\bar A\in \R^{2\times 2}$. The minimum is achieved for $b^{\rm min}_{j}=-\bsc^{-1}_{ji}(\C\bar A)_{i3}$, and
$$(\C(\bar A+b^{\rm min}\odot e_3))_{i3}=0.$$
Also
$$\hat \C\bar A\cdot \bar A\ge c_\C |\bar A|^2,$$
for every symmetric matrix $\bar A\in \R^{2\times 2}$.
\end{lemma}
\proof
The statements concerning $\bar \C$ follow by an easy computation.
To prove the statement concerning $\hat\C$ note that
$$
\int_{-1/2}^{1/2}\bar \C (\bar B +x_3\bar A)\cdot(\bar B +x_3\bar A)\,dx_3\ge c_\C\int_{-1/2}^{1/2}|\bar B +x_3\bar A|^2\,dx_3\ge \frac{c_\C}{12}|\bar A|^2,
$$
and since
$$
\int_{-1/2}^{1/2}\bar \C (\bar B +x_3\bar A)\cdot(\bar B +x_3\bar A)\,dx_3=\bar\C^{(0)}\bar B\cdot \bar B+
2 \bar\C^{(1)}\bar A\cdot \bar B+\bar\C^{(2)}\bar A\cdot \bar A,
$$
we have that
$$
\hat \C\bar A\cdot \bar A=12 \min_{\bar B\in \R^{2\times2}_{\rm sym}}\int_{-1/2}^{1/2}\bar \C (\bar B +x_3\bar A)\cdot(\bar B +x_3\bar A)\,dx_3.
$$
\QED

Thus, from \eqref{ZN1} and \eqref{ZM1}, and using \eqref{A4}, \eqref{A4b},  \eqref{A4t} and \eqref{A7}, we deduce \eqref{A5}.
Hence from \eqref{A5} we find $\bar Z$ and from \eqref{zj} we find $z$. Thus also $Z$ is completely known and hence, from the relation, $\C Z=S^\calL+\Lambda$, also $\Lambda$ is known in terms of $S^\calL$.

We now compute $f^\perp(S^\calL)$. We have
\begin{align*}
f^\perp(S^\calL)&=\calF^\perp(S^\calL,\Lambda)=\calF^\perp(S^\calL,\C Z-S^\calL)\\
&=\frac 12 \int_{\Om} \C Z\cdot  Z-\C^{-1} S^\calL\cdot  S^\calL\,d x.
\end{align*}
Let us write \eqref{zj} as follows
$$z=z^{\rm min}+\ssf\quad\mbox{with }z^{\rm min}_j:=-\bsc^{-1}_{ji}(\C\bar Z)_{i3},\quad\ssf_j:=\bsc^{-1}_{ji}F_{i3},$$
then 
\begin{align*}
 \C Z\cdot  Z&= \C (\bar Z+z\odot e_3)\cdot(\bar  Z+z\odot e_3)\\
 &=\C (\bar Z+z^{\rm min}\odot e_3)\cdot(\bar  Z+z^{\rm min}\odot e_3)\\
 &\hspace{1cm}+2\C (\bar Z+z^{\rm min}\odot e_3)\cdot\ssf\odot e_3+\C\, \ssf\odot e_3\cdot\ssf\odot e_3\\
 &=\bar\C \bar Z\cdot\bar  Z+\bsc\, \ssf\cdot\ssf,
 \end{align*}
 where to obtain the last equality we have used Lemma \ref{positive}.
Thus
$$
f^\perp(S^\calL)=\frac 12 \int_{\Om} \bar\C \bar Z\cdot \bar Z-\C^{-1} S^\calL\cdot  S^\calL+\bsc\, \ssf\cdot\ssf\,d x,
$$
which is equivalent to \eqref{A8}.

\vspace{5mm}

\noindent
{\bf Acknowledgement.} We would like to thank an anonymous referee for pointing out a gap in the proof of the existence of the recovery sequence present
in a previous version of our paper.


\end{document}